\documentclass[sigconf,nonacm,pbalance]{acmart}

\usepackage{enumitem}
\usepackage{algorithm}
\usepackage[noend]{algpseudocode}
\usepackage{placeins}

\AtBeginDocument{%
  }
\renewcommand\footnotetextcopyrightpermission[1]{}
\title[Odin]{Odin: Primitive-Level Synchronization for Distributed Point-Based Neural Rendering}

\author{Zhenxiang Ma}
\affiliation{
  \institution{Shanghai Jiao Tong University}
  \city{Shanghai}
  \country{China}
}
\affiliation{
  \institution{Shanghai AI Laboratory}
  \city{Shanghai}
  \country{China}
}
\email{mazhenxiang@sjtu.edu.cn}

\author{Zeyu He}
\affiliation{
  \institution{Shanghai AI Laboratory}
  \city{Shanghai}
  \country{China}
}
\email{hezeyu@pjlab.org.cn}

\author{Yuanzhen Zhou}
\affiliation{
  \institution{Shanghai AI Laboratory}
  \city{Shanghai}
  \country{China}
}
\email{zhouyuanzhen@pjlab.org.cn}

\author{Zhenyu Yang}
\affiliation{
  \institution{Shanghai AI Laboratory}
  \city{Shanghai}
  \country{China}
}
\email{yangzhenyu@pjlab.org.cn}

\author{Yuchang Zhang}
\affiliation{
  \institution{Shanghai AI Laboratory}
  \city{Shanghai}
  \country{China}
}
\email{zhangyuchang@pjlab.org.cn}

\author{Miao Tao}
\affiliation{
  \institution{Shanghai AI Laboratory}
  \city{Shanghai}
  \country{China}
}
\email{taomiao@pjlab.org.cn}

\author{Rong Fu}
\affiliation{
  \institution{Shanghai AI Laboratory}
  \city{Shanghai}
  \country{China}
}
\email{furong@pjlab.org.cn}

\author{Jidong Zhai}
\affiliation{
  \institution{Tsinghua University}
  \city{Beijing}
  \country{China}
}
\email{zhaijidong@tsinghua.edu.cn}

\author{Hengjie Li}
\authornote{Corresponding author.}
\affiliation{
  \institution{Shanghai AI Laboratory}
  \city{Shanghai}
  \country{China}
}
\affiliation{
  \institution{Shanghai Innovation Institute}
  \city{Shanghai}
  \country{China}
}
\email{lihengjie@pjlab.org.cn}

\renewcommand{\shortauthors}{Ma et al.}

\begin{document}

\begin{abstract}
Point-based neural rendering (PBNR) represents 3D scenes as explicit, trainable primitives and has become an important foundation for high-quality reconstruction and emerging embodied-AI and world-model pipelines. Unlike layer-structured neural networks, PBNR exposes primitive-indexed dependencies: each view observes and updates only a sparse, view-dependent subset of the mutable scene state. As large scenes push PBNR toward distributed training and optimized renderers reduce per-view computation, global task- or iteration-level barriers increasingly place synchronization, rather than rendering, on the critical path.

Odin breaks these barriers with primitive-level synchronization. Its ahead-of-time scheduler uses stable PBNR locality and phase order to place low-conflict overlap windows, while the runtime validates primitive publication before later state observation and publishes the pending updates that later work may observe. Odin provides a quality-first path that preserves synchronized-training visibility and a throughput-first path that uses overlap and gradient evidence to admit only small, low-impact delayed reads; structural changes and high-impact cases remain synchronized.

Across four existing PBNR pipelines and 13 non-city 8-GPU scenes, Odin improves throughput by 1.22$\times$ on average and hides 82\% of critical-path wait while preserving reconstruction quality. In a MatrixCity mixed-parallel case study up to 64 GPUs, Odin improves over Grendel by up to 1.89$\times$ without changing renderer kernels, optimizers, training budgets, or model capacity.
\end{abstract}


\maketitle

\section{Introduction}
\label{sec:intro}
Point-based neural rendering (PBNR), represented by 3D Gaussian Splatting (3DGS)~\cite{3DGS} and follow-up methods~\cite{2DGS,minisplat,Eagles,compact,stopthepop}, is becoming a foundational algorithmic family for embodied intelligence and world models~\cite{abou2024physically,lu2025gwmscalablegaussianworld}. It reconstructs explicit, trainable 3D scene state from multi-view observations and trains each camera view by rendering, backpropagating, and updating only visible primitives. At the systems level, PBNR follows a tensor-training execution model: GPU kernels operate over aligned primitive-attribute and optimizer-state tensors. Semantically, each view only reads and updates a sparse, view-dependent subset of primitive indices. This creates a systems mismatch absent from regular layer-by-layer neural networks: the dependency unit is a primitive index, but current distributed PBNR commonly publishes updates through global task- or iteration-level barriers. Consequently, a later view can be forced to wait for primitive updates outside its observable scope.

\begin{figure}[t]
\centering
\includegraphics[width=1.0\columnwidth]{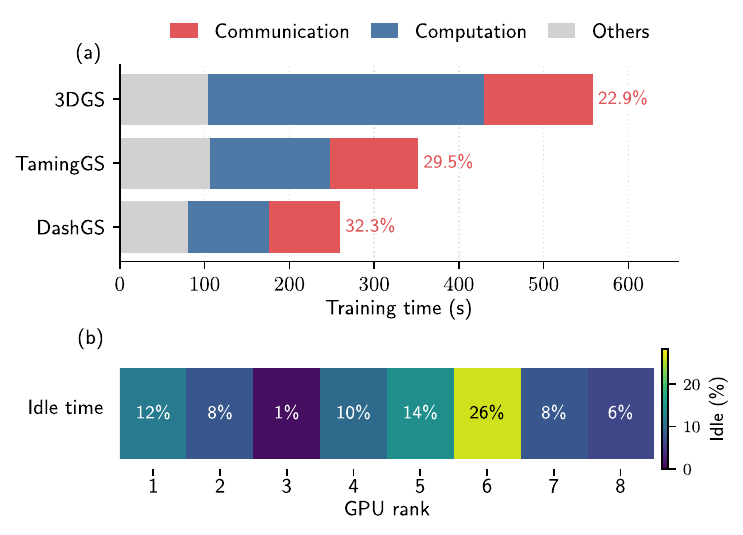}
\caption{Motivating synchronization overhead in distributed point-based neural rendering (PBNR). Top: optimized variants expose a growing communication fraction as rendering and update computation shrink. Bottom: per-GPU idle time in an 8-GPU run shows that a global barrier turns rank-local skew into system-wide waiting.}
\label{fig:motivation0}
\end{figure}

Recent PBNR optimizations make this mismatch increasingly costly. They shorten rendering and update phases while leaving synchronization exposure largely intact. Coarse barriers have always existed in distributed PBNR, but earlier computation-heavy pipelines could hide more of their cost. Figure~\ref{fig:motivation0} shows this trend: exposed communication rises from 22.9\% in 3DGS to 29.5\% in TamingGS and 32.3\% in DashGS. In an 8-GPU run, per-GPU idle time also ranges from 1\% to 26\%, indicating that coarse barriers amplify imbalance across workers. The key systems question is therefore not only how much state is exchanged, but which primitive updates actually need to block a later view.

Existing systems optimize where PBNR state and views run, but not when a primitive update becomes visible to later views. Distributed PBNR systems exploit locality for ownership, placement, partitioning, sparse exchange, or load balance~\cite{Grendel, cityGS, zhao2025scalingpointbaseddifferentiablerendering}; Gaian~\cite{zhao2025scalingpointbaseddifferentiablerendering}, the closest locality-based system, uses point-based differentiable-rendering access patterns for point placement and image-to-GPU assignment. These systems decide where primitives and views run and how much state moves, but their remaining exchange still becomes visible through coarse task- or iteration-level barriers. General distributed deep-learning schedulers operate at layer, tensor, stage, or collective boundaries, while relaxed-gradient methods have mainly been validated for dense neural networks or tensor-averaging regimes. PBNR has no deep layer hierarchy, and its relevant conflicts sit below the tensor boundary: two tasks with identical phase order can be independent or conflicting solely because their views activate different primitive indices.

The challenge is that the activated primitives that determine whether a barrier is necessary are irregular, dynamic, and implicit at scheduling time. They are irregular: activation depends on geometry, opacity, visibility, and numerical state. They are dynamic: densification, pruning, and reset change the primitive set during training. They are also implicit: the phase task graph exposes only legal phase order, while tensor-level execution hides primitive-index dependencies. Early geometry and co-visibility are available in time to schedule, but they overestimate conflicts. Realized activity and gradient impact are more faithful, but they arrive after the overlap window must already have been placed. A useful system must therefore schedule tentative overlap early, then validate primitive publication before state observation.

We break global barriers with static prediction and runtime correction over primitive publication. The design first builds a relative locality graph (RLG) from stable co-visibility, then schedules task-unit order and places overlap windows between communication and computation jointly with the phase task graph. Runtime execution uses \emph{Shadow Graph} to give overlapped units versioned logical views over shared physical state. Before a later task observes state, the runtime validates the predicted window and publishes only the primitive updates that task may observe. When a prediction is unsafe, it falls back to primitive-level waiting rather than rollback. We implement this design as Odin, a distributed PBNR training system.

Odin supports two paths. The quality-first path is equivalent to synchronized training under the scheduled order: it removes only waits whose primitive scopes are validated as disjoint, or whose observable updates have already been published. The throughput-first path targets deployment workloads and capture patterns where small, low-impact RAW overlaps block larger overlap windows. This is PBNR-specific rather than generic asynchronous training: compositing, occlusion, and transmittance give primitive interactions physical and numerical weights, making some dominant and others weak~\cite{occluGS,ODAGS}. Odin admits only delayed reads with both small delayed scope and small producer gradients; writes, structural changes, missing evidence, and high-impact cases stay synchronized. Across four PBNR pipelines and 13 non-city 8-GPU scenes, Odin improves throughput by 1.22$\times$ on average; in a MatrixCity mixed-parallel case study up to 64 GPUs, it reaches 1.89$\times$ over Grendel.

In summary, this paper makes the following contributions:

\begin{itemize}

\item We identify a synchronization-boundary mismatch in distributed PBNR: optimized renderers expose global barriers, while view dependencies remain sparse at primitive-index granularity (Sections~\ref{sec:intro}--\ref{sec:overview}).

\item We introduce primitive-level synchronization to break these barriers. Odin publishes only primitive updates that later views may observe, replacing whole-task or whole-iteration waits with validated primitive publication (Sections~\ref{sec:method}--\ref{subsec:runtime}).

\item We present a two-graph scheduler/runtime co-design. The AOT scheduler augments the phase task graph with RLG to expose overlap missed by phase order; runtime validation corrects unsafe windows, while \emph{Shadow Graph} stages overlap without full replication (Sections~\ref{sec:method}--\ref{sec:implementation}).

\item We implement Odin in four PBNR pipelines without changing kernels, optimizers, budgets, or model capacity; evaluation shows 1.22$\times$ average speedup, up to 1.89$\times$ over Grendel, stable quality, and ablations that attribute gains to primitive-scoped synchronization (Section~\ref{sec:evaluation}).

\end{itemize}

\section{Background}
\label{prem}

\subsection{Point-Based Neural Rendering}
\label{subsec:pbnr}

\FloatBarrier

\begin{figure}[t]
  \centering
  \includegraphics[width=1.0\columnwidth]{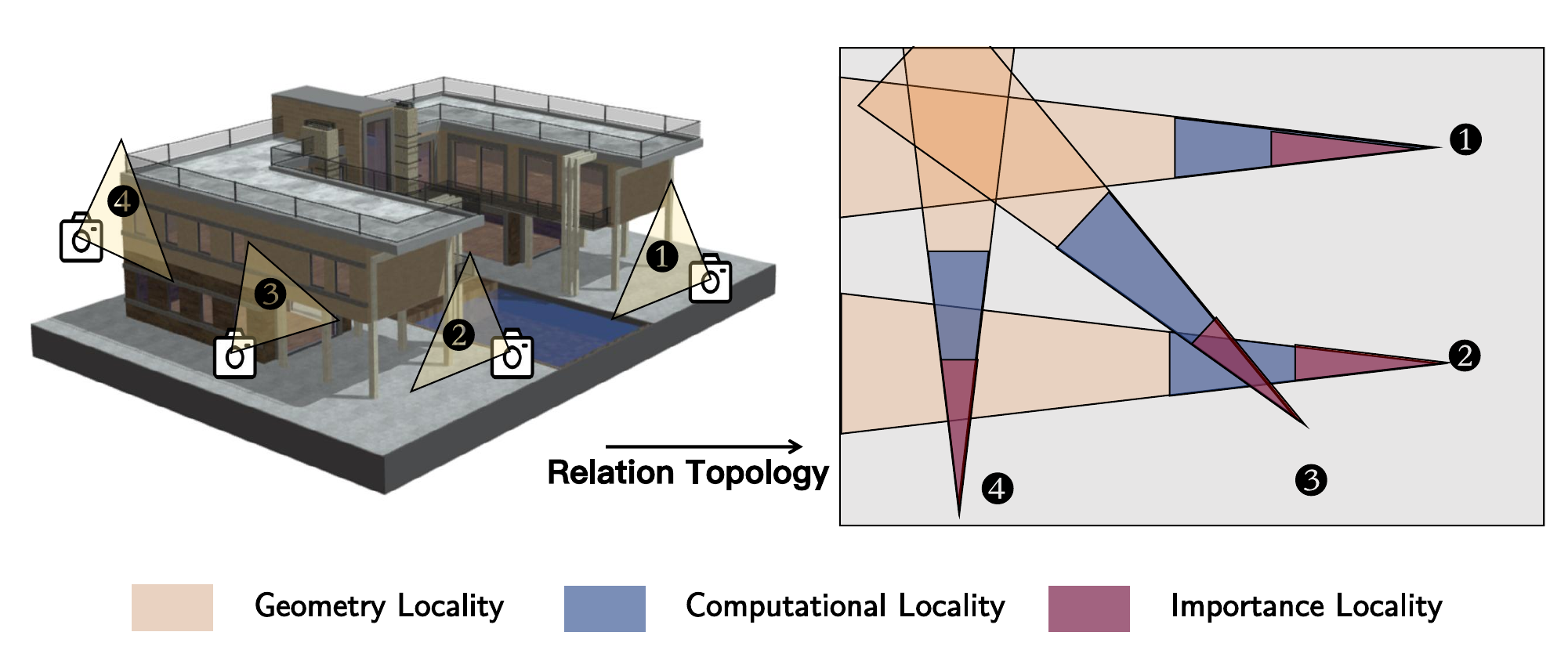}
  \caption{Running four-view locality example. \textbf{Left}: four numbered training views over the same primitive pool. \textbf{Right}: beige, blue, and purple show geometry locality, computational locality, and importance locality. They are progressively later and more selective evidence: candidate primitives a view may touch, primitives it actually activates, and active primitives with small measured impact.}
  \label{fig:locality}
\end{figure}

Figure~\ref{fig:locality} gives the semantic abstraction used throughout the paper. A PBNR model is not a deep stack of layers; it is a mutable pool of explicit point-like scene primitives, denoted by \(\Theta\). Each primitive contains geometry, opacity, appearance parameters, and optimizer state. Implementations store these fields as aligned tensors for GPU efficiency, but the dependency unit remains the primitive index across those tensors. Each numbered item on the left of the figure is a training view: it renders one camera, computes an image loss, and updates only the primitives that participate in that view. The systems opportunity is visible immediately: although the primitive pool is global, a single view usually touches only a localized subset.

The right side of Figure~\ref{fig:locality} shows the three locality signals that Odin uses. They are not three separate correctness models; they are the same view--primitive relation observed at different times. \emph{Geometry locality} is the earliest signal. Before rendering, camera geometry or owner-region metadata gives a conservative candidate set of primitives a view may touch. It is available in time for scheduling, but it can include primitives that later contribute little or not at all. \emph{Computational locality} is the realized active set after rendering and backpropagation. It is more faithful, because it records which primitives actually participate in the computation, but it is only fully known after the view has executed. \emph{Importance locality} attaches gradient evidence to the realized active set. It captures the PBNR-specific observation that some realized interactions are numerically small, which creates the throughput-first opportunity used later in the paper.

This timing is the central difficulty. If a scheduler waits for realized activity and gradients, the overlap window has already passed. If it trusts only early geometry, it often keeps unnecessary barriers. Odin therefore uses these signals according to when they appear: stable early locality guides scheduling, conservative scopes validate the quality-first path before state observation, and importance evidence gates bounded RAW delayed reads in the throughput-first path. Structural changes such as densification, pruning, and opacity reset still affect the primitive pool globally and remain barrier-scoped.

The synchronization checks later in the paper rely on three levels of view--primitive scope: the full primitive pool, a conservative pre-observation scope, and the realized active set. For a view \(i\), these scopes follow the nesting shown in Figure~\ref{fig:locality}:
\[
\Theta \supseteq \mathcal{G}_i \supseteq \mathcal{C}_i .
\]
\(\Theta\) is the full mutable primitive pool. \(\mathcal{G}_i\) is the conservative candidate scope available before the view reads mutable state. \(\mathcal{C}_i\) is the realized active subset observed after execution. Figure~\ref{fig:locality} can therefore be read as a simple rule: beige is early but conservative, blue is faithful but late, and purple adds measured impact.

A task unit is a small batch of training views. We call the first point at which a task reads mutable primitive parameters or optimizer state its state-observation point. For a single view, \(R_i^+\) and \(W_i^+\) are conservative read and update scopes used before that point. The \(+\) superscript means an admission upper bound, not the exact set that will execute. For the render-loss phases evaluated here, \(W_i^+\subseteq R_i^+\subseteq\mathcal{G}_i\); pipelines with broader update effects declare a wider \(W_i^+\) or synchronize the phase.

For task unit \(U\), Odin takes the union over its views: \(R^+(U)=\cup_{i\in U}R_i^+\), \(W^+(U)=\cup_{i\in U}W_i^+\), and \(\mathcal{C}_U=\cup_{i\in U}\mathcal{C}_i\). For an active primitive \(p\), \(g_U(p)\) is the accumulated per-primitive gradient magnitude after the pipeline's normal gradient scaling. These post-execution signals support later refinement and importance-aware admission; the early scopes support scheduling and quality-first validation.

\begin{table}[t]
\centering
\caption{Key primitive-scope notation used by Odin.}
\label{tab:locality_terms}
\small
\setlength{\tabcolsep}{4pt}
\renewcommand{\arraystretch}{1.05}
\begin{tabular}{@{}p{0.18\columnwidth}p{0.70\columnwidth}@{}}
\toprule
Symbol & Meaning \\
\midrule
\(\Theta\) & Mutable primitives in the scene. \\
\(\mathcal{G}_i\) & Geometry locality: candidate primitive scope before state observation. \\
\(\mathcal{C}_i\) & Computational locality: realized active primitive set after execution. \\
\(R_i^+, W_i^+\) & Conservative read/update scopes used for admission. \\
\(g_U(p)\) & Gradient evidence for active primitive \(p\). \\
\bottomrule
\end{tabular}
\end{table}

\begin{figure}[t]
  \centering
  \includegraphics[width=1.0\columnwidth]{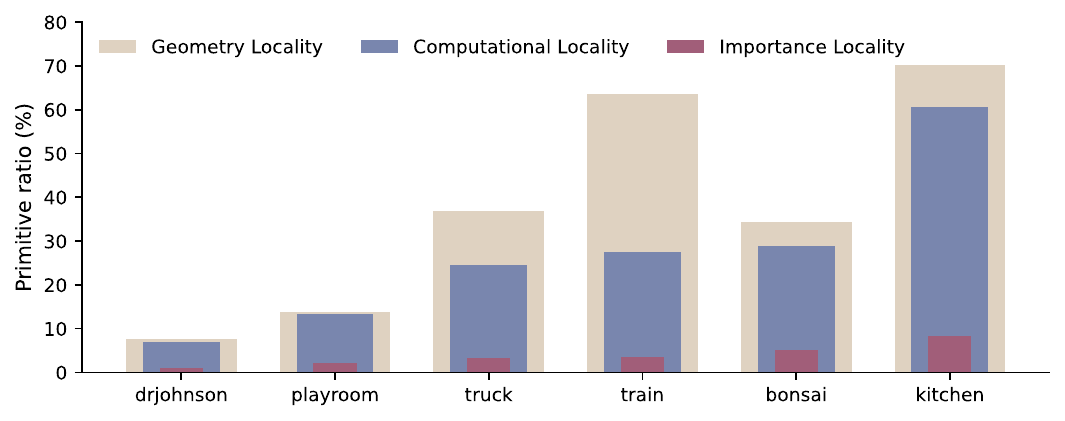}
  \caption{Locality-scope ratios in real scenes after 7{,}000 iterations. Each bar reports the primitive fraction retained by one view under geometry locality, computational locality, or importance locality. The drop from geometry to realized activity shows why early scopes are conservative; the further drop after importance filtering shows why bounded low-impact RAW delayed reads can expose additional throughput opportunity.}
  \label{fig:locality_ratio}
\end{figure}

Real scenes show the same pattern at scale. Figure~\ref{fig:locality_ratio} reports that even conservative geometry-locality scopes are smaller than the full primitive pool, while scopes from computational and importance locality are smaller still. The gap creates scheduling opportunity, and the scene-to-scene variation explains why Odin combines ahead-of-time scheduling with runtime validation.

\subsection{Distributed PBNR}
\label{subsec:dist_pbnr}

Distributed PBNR usually scales training through data parallelism (DP) or partitioned-state execution such as mixed parallelism (MP)~\cite{Grendel}. DP replicates \(\Theta\) and reconciles updates; MP shards \(\Theta\) and communicates view-needed state. Both reduce memory pressure or communication volume, but typically keep task- or iteration-level synchronization.

\begin{figure}[t]
  \centering
  \includegraphics[width=1.0\columnwidth]{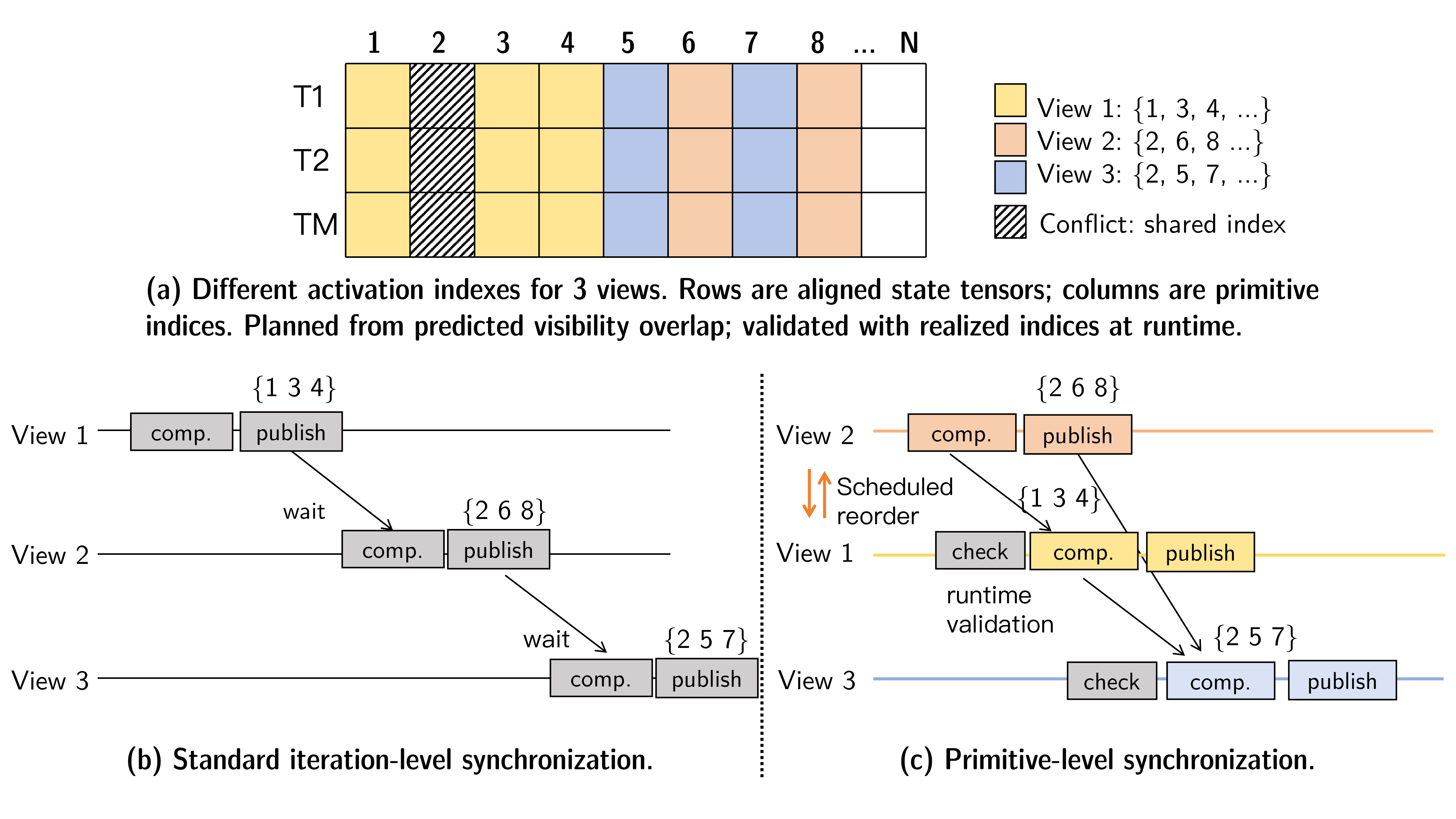}
  \caption{Primitive-index synchronization over aligned tensor state. Colors mark view scopes and hatching marks shared indices; the few columns shown are schematic, while real PBNR scenes typically have \(N\) in the millions. Panels (b) and (c) emphasize dependency and publication order rather than proportional execution cost; block lengths are schematic. Standard iteration-level synchronization waits after each publication. Odin schedules a low-conflict order, overlaps disjoint scopes, and validates primitive publication before later state reads; shared pending indices wait for publication.}
  \label{fig:locality_execution}
\end{figure}

Figure~\ref{fig:locality_execution} translates the locality abstraction into the program state seen by a distributed runtime. In memory, PBNR state is stored as aligned tensors; semantically, each column across those tensors is one primitive index. This makes the mismatch independent of whether execution uses DP replication or MP sharding: the physical state may be tensor-shaped, but visibility conflicts are primitive-index scoped. A standard barrier serializes views even when their primitive-index scopes are disjoint. Odin instead schedules low-conflict views to expose overlap and validates primitive publication before state reads, so only shared pending primitive updates delay later work.

\section{Motivation and Approach}
\label{sec:overview}
Figures~\ref{fig:locality} and~\ref{fig:locality_execution} expose both the opportunity and the constraint. The locality example shows that views usually touch only a subset of the global primitive pool. The tensor-index example shows why this subset must become a synchronization boundary: a global barrier makes a later view wait even when primitive-index scopes are disjoint, but removing synchronization entirely would let a view read shared pending indices. The system problem is therefore to replace an iteration-wide boundary with scheduled primitive-level synchronization checks over only the updates a later view may observe.

Odin follows a predict-and-validate principle because exact primitive-index conflicts are unavailable when the overlap window must be placed. A purely ahead-of-time scheduler can use early geometry, but that signal is conservative; a purely reactive scheduler observes faithful primitive activity only after the overlap opportunity has passed. Odin therefore predicts task-unit order and overlap windows from a stable locality prior and phase order, then validates each planned window before state observation. Failed predictions trigger primitive-scoped synchronization on the affected updates, while realized activity and gradients refine later dispatch and admission decisions.

Primitive scopes tell Odin when a later task can run without observing conflicting unpublished updates; global structural events remain synchronized. This is the quality-first path: it preserves synchronized-training visibility while replacing a global barrier with primitive-level waits. In practical distributed PBNR runs, however, a small read-after-write overlap can still block an otherwise useful window. The throughput-first path addresses this second bottleneck with a delayed-scope ratio and producer-gradient gate, admitting only RAW delayed reads whose PBNR rendering signals indicate small numerical effect. Odin therefore first narrows synchronization to primitive scopes, then uses importance to broaden that boundary where the workload supports it, rather than replacing training with generic asynchronous execution.

\section{Method}
\label{sec:method}
Odin's method is organized into three parts: graph construction, graph scheduling, and graph execution. Graph construction builds the relative locality graph (RLG) and binds it to the phase task graph, giving Odin both a data-relation signal and the legal phase order of the PBNR pipeline. Graph scheduling performs the ahead-of-time optimization: static data scheduling arranges task-unit order, and static asynchronous scheduling places candidate overlap windows on the phase task graph. Graph execution is the dynamic counterpart for shared tensor state: it executes overlapped units through \emph{Shadow Graph}, refines ready-unit dispatch through dynamic data scheduling, and uses dynamic asynchronous scheduling to revalidate planned windows before state observation.

\subsection{Graph Construction}
\label{subsec:graph_build}

The first stage constructs the graph abstraction over which Odin schedules. Odin separates opportunity discovery from visibility validation: the static graph ranks promising view-pair overlaps, and graph execution validates observable primitive updates before state is consumed. This lets the RLG use a strong, stable locality prior without fixing absolute primitive dependencies before training.
\begin{figure}[t]
  \centering
  \includegraphics[width=1.0\columnwidth]{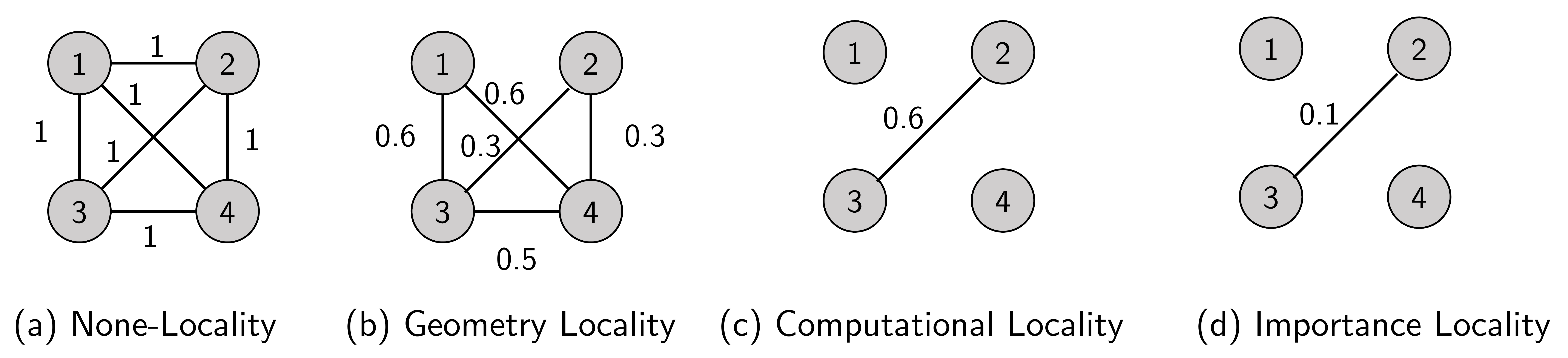}
  \caption{Relative locality graph (RLG) for the four-view example. Nodes are views and edges are predicted coupling. The panels compare no graph, geometry-, computation-, and importance-based RLGs. Odin builds the ahead-of-time RLG from stable co-visibility and uses later evidence to refine dispatch or admission.}
  \label{fig:RLG}
\end{figure}

\subsubsection{Relative locality graph}
\label{subsubsec:data_graph}

The core abstraction is a \emph{relative locality graph} (RLG), a weighted undirected graph
\[
\mathcal{R} = (\mathcal{V}, \mathcal{E}, w),
\]
where each vertex \(v_i \in \mathcal{V}\) represents a training data item, instantiated as an input view in PBNR; \(\mathcal{E}\) stores retained locality edges; and \(w_{ij}\) is a relative coupling score between views \(i\) and \(j\). Larger weights mean stronger expected contention. The RLG is an additional scheduling input outside the phase task graph and renderer autograd graph: it ranks overlap opportunities with stable, updateable locality evidence, while runtime validates visibility. This suits PBNR because primitive positions and update regions move during training; view-pair coupling is more stable than sampling-space indices such as octrees. For scheduling scores, a missing edge uses \(w_{ij}=0\) and is validated at runtime, not treated as proven independence.

Figure~\ref{fig:RLG} shows why RLG construction is a design choice rather than a direct lookup. Coarse signals yield dense graphs and few schedulable opportunities; faithful signals appear too late or change too quickly to serve as the initial RLG.

\paragraph{\textbf{Geometry-based RLG}}
A geometry-based RLG from projection overlap or owner-region metadata is cheap, early, and useful for conservative scopes. It is a natural fit for top-down city-scale captures, where cross-view coupling is often weaker and spatial coverage is more separable. For general PBNR, however, it is too coarse: overlapping projected regions may correspond to different surfaces, occlusions, or weakly shared primitives, making the graph dense and discarding many useful overlap windows.

\paragraph{\textbf{Computation-based RLG}}
A computation-based RLG follows an inspector--executor pattern: first trace or pre-run visibility and primitive accesses, then schedule against the materialized dependency graph. It is faithful, and Odin uses realized activity after execution for summaries and dispatch refinement. As the ahead-of-time RLG, however, it is impractical: tracing is expensive, arrives after the useful overlap window, and becomes stale as primitives move, split, or disappear.

\paragraph{\textbf{Importance-based RLG}}
It should rank not only whether two views overlap, but whether their overlap is likely to matter. Gradients provide one numeric signal: compositing, occlusion, and transmittance make some nonzero primitive overlaps dominant and others weak~\cite{occluGS,ODAGS}. This post-backward evidence supports delayed-scope and gradient gates in Graph Execution, but it cannot place the initial overlap window.

Odin instead builds the ahead-of-time RLG from a physical importance prior: stable co-visibility. We use SfM tracks~\cite{colmap, vggt}: observations that repeatedly share tracks are likely to access related scene content, while observations without shared tracks are less likely to do so. This gives the scheduler an importance-like ranking before gradients exist, more selective than coarse geometry and more stable than a traced primitive-access graph.

Let \(\mathcal{P}_i\) denote the set of SfM tracks associated with data item \(i\). If \(\mathcal{P}_i\cup\mathcal{P}_j\) is nonempty, Odin defines the RLG edge weight by Jaccard similarity:
\begin{equation}
    w_{ij} = \frac{|\mathcal{P}_i \cap \mathcal{P}_j|}{|\mathcal{P}_i \cup \mathcal{P}_j|}.
    \label{eq:jaccard}
\end{equation}
If the union is empty, Odin sets \(w_{ij}=0\) and relies on runtime validation rather than treating the pair as proven independent. For task-unit scheduling, Odin summarizes two units by the largest retained view-pair coupling between them. We compute shared-track counts through an inverse track-to-item mapping, cap unusually long tracks to limit outliers, and retain only the top-\(k\) neighbors per item, where lowercase \(k\) is the neighbor-retention parameter and uppercase \(K\) later denotes the number of logical regions. These choices affect only ranking; publication safety is checked at runtime. Table~\ref{tab:mapping_overhead} reports the overhead.

SfM-based construction is one RLG instantiation; later realized activity and gradient evidence can refresh transition scores. Outdated edges reduce scheduling quality rather than quality-first safety, because Graph Execution validates planned overlap before state observation.

\subsubsection{Phase task graph}
\label{subsubsec:task_graph}
In addition to the RLG, Odin uses the existing phase task graph of a PBNR training pipeline. This graph records legal phase order, such as scope extraction, preprocessing, rendering, backward computation, communication, optimizer update, and required barriers. It is not the tensor-level autograd graph inside the renderer; it is the coarse execution scaffold that existing systems can schedule. The phase task graph constrains legal execution but lacks the data-relation signal needed to identify low-conflict view transitions. Odin schedules on both graphs: the phase task graph preserves legal execution order, while the RLG supplies that missing primitive-state coupling signal.

\subsection{Graph Scheduling}
\label{subsec:graph_schedule}
 \begin{figure}[t]
    \centering
    \includegraphics[width=1.0\columnwidth]{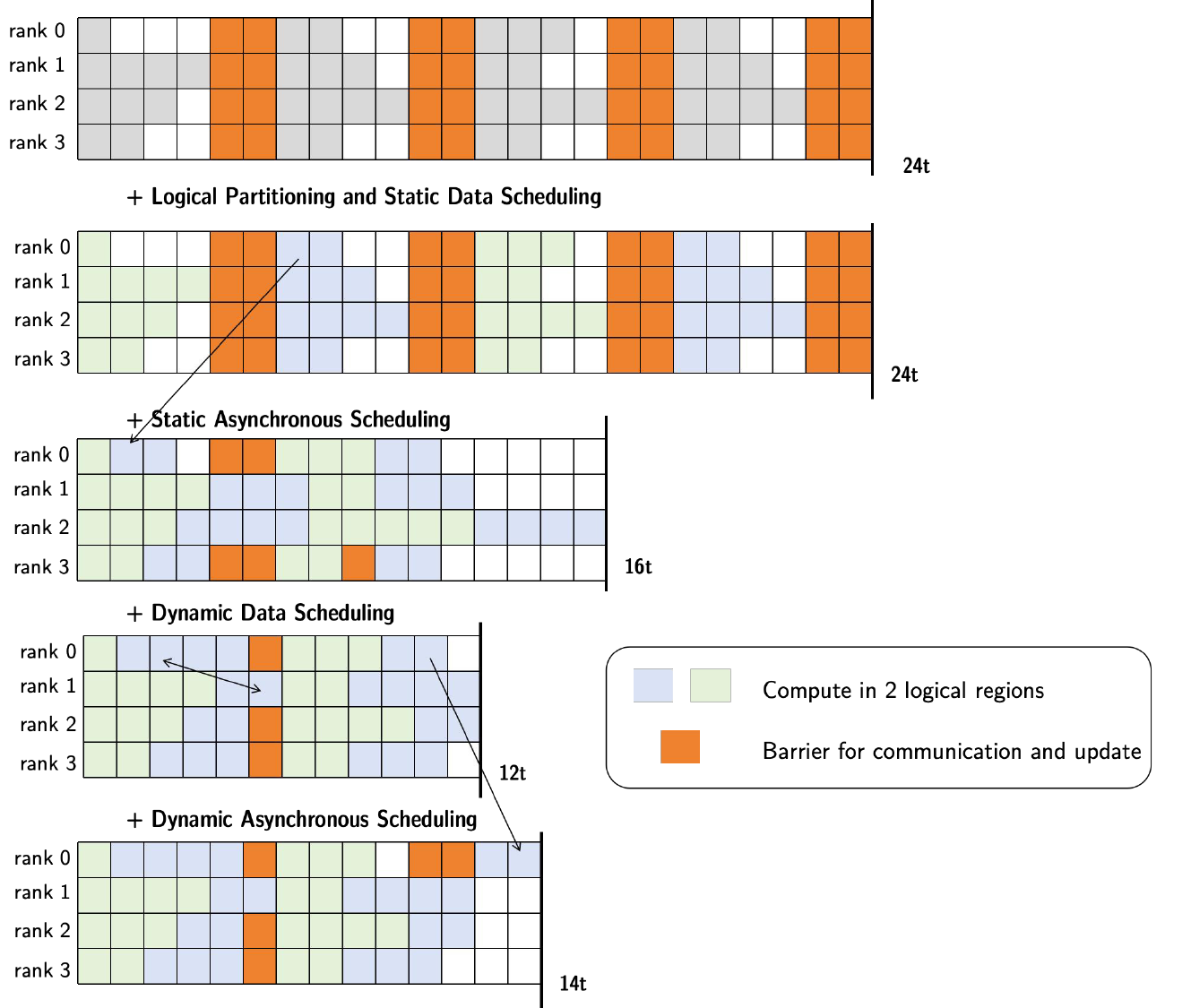}
    \caption{Odin scheduling pipeline on four ranks. Static scheduling uses the RLG prior and phase order to arrange task units and place candidate overlap windows; runtime execution adjusts dispatch and validates admission before state observation. Blue/green cells are computation from two logical regions, orange cells are communication and optimizer-update barriers where publication may occur, and white gaps are idle bubbles.}
    \label{fig:pipeline}
\end{figure}

Given the RLG and the phase task graph, Odin compiles locality into a scheduled graph in three steps: logical partitioning, static data scheduling, and static asynchronous scheduling. Figure~\ref{fig:pipeline} shows the intended transformation: the original phase order contains repeated communication/update barriers, LP and SDS reorder work so that weakly coupled regions appear near overlap opportunities, and SAS marks where communication can be launched ahead of a later validation point. The output records region assignment, task-unit order, candidate overlap windows, and validation points. It is not a final visibility decision: graph execution later dispatches ready work and validates, refines, or delays each planned window before state observation.

\paragraph{\textbf{Logical Partitioning (LP)}}
LP partitions the RLG into \(K\) logical groups: scheduling bins, not state shards or device owners. A good partition balances groups, keeps strong coupling within groups, and leaves weak edges across groups; \(K\) controls the granularity of SDS's inter-group rotation. The planner also keeps a region-level coupling summary so later steps can choose weakly coupled region transitions before selecting individual views. We use lightweight balanced clustering~\cite{mcqueen1967some,least}; other clustering choices can replace it without changing Odin's publication rules.

\paragraph{\textbf{Static Data Scheduling (SDS)}}
SDS rotates over the \(K\) groups to build task-unit order, preserving the epoch workload, task-unit size, and phase precedence while exposing low-conflict transitions. Odin first chooses a traversal over logical groups using the region-level coupling summary, preferring weakly coupled successors while keeping groups balanced. It then instantiates this traversal into concrete task units: each new unit is filled from the selected group by greedy local selection against the preceding scheduled unit \(U_{\mathrm{prev}}\). A candidate item \(v\) is scored by
\begin{equation}
s(v\mid U_{\mathrm{prev}})=\sum_{u\in U_{\mathrm{prev}}} w_{uv}.
\label{eq:incremental_cost}
\end{equation}
SDS repeatedly selects low-score candidates until the unit is filled. It does not need to make every adjacent pair independent; it only increases the number of transitions worth attempting. If an ordering prediction is wrong, Graph Execution synchronizes the affected primitive updates before the later task observes state.

\paragraph{\textbf{Static Asynchronous Scheduling (SAS)}}
Using the \(K\)-group SDS order, SAS maps task-unit transitions onto PBNR phases instead of treating each unit as an indivisible job. Each phase operator is bound to a task-unit id, producing events such as preprocessing, rendering, backward computation, communication, and optimizer update for \(A\) and \(B\). For adjacent \(A\rightarrow B\), a global barrier waits for \(A\)'s backward-generated primitive updates to communicate and publish before \(B\)'s state observation. SAS statically marks this edge as blocking or as a candidate overlap window: candidates launch \(A\)'s publication communication asynchronously and place a validation point before \(B\) observes state; high RLG coupling, structural phases, or windows whose profiled live state exceeds the memory budget stay blocking. The decision combines RLG coupling with a lightweight phase profile of coarse render, backward, communication, and update durations. If the planned window is shorter than the actual publication time, Graph Execution leaves a residual wait, shown as white bubbles in Figure~\ref{fig:pipeline}. SAS never admits reads by itself; Graph Execution validates primitive scopes at the marked point.

\subsection{Graph Execution}
\label{subsec:runtime}
\begin{figure}[t]
  \centering
  \includegraphics[width=1.0\columnwidth]{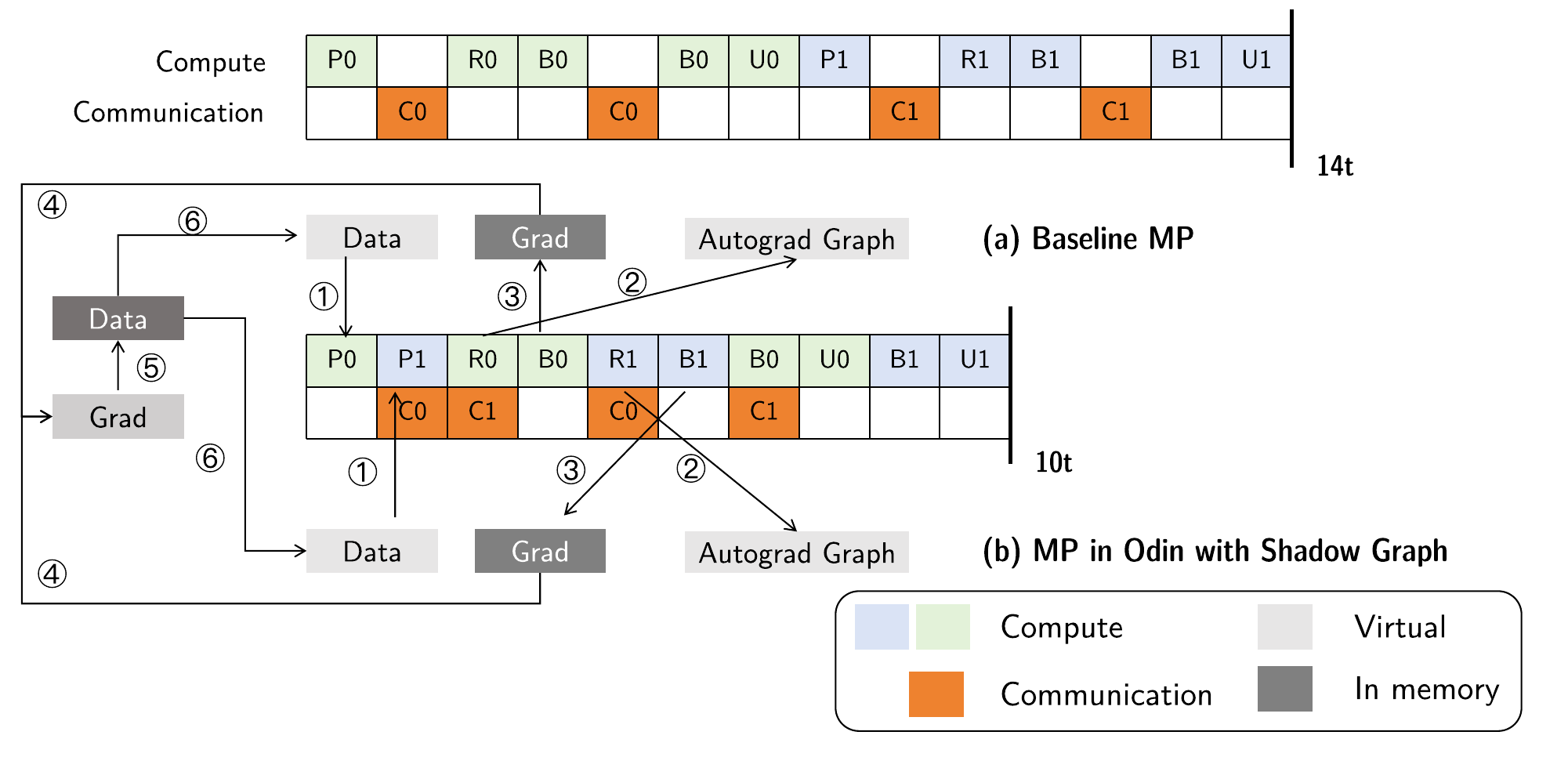}
  \caption{\emph{Shadow Graph} execution for overlapped task units. Phase labels \(P/R/B/U/C\) denote preprocessing, rendering, backward computation, optimizer update, and communication; suffixes 0/1 identify two task units. Each unit uses private logical views for local execution and staged gradient publication while physical model state remains shared. Numbered arrows summarize private read, local autograd, staged update, staged publication after validation, shared-state refresh, and reuse.}
  \label{fig:shadow_graph}
\end{figure}

Graph execution realizes the scheduled graph produced by graph scheduling. The scheduled graph fixes task-unit order and candidate overlap windows, but runtime behavior determines which ready unit should run next and whether a planned window is safe before state observation. Odin therefore executes the plan with three coupled mechanisms: \emph{Shadow Graph} supports overlapped execution without full model replication, dynamic data scheduling refines ready-unit dispatch under observed load, and dynamic asynchronous scheduling validates or corrects planned overlap using primitive scopes and realized evidence.

\paragraph{\textbf{\emph{Shadow Graph} execution}}
A direct execution issue arises once Odin overlaps phase-level work across task units. The synchronization boundary is primitive-index scoped, but the physical model state remains a set of shared aligned tensors and training loops normally expose one serialized producer--consumer chain over that state. If a later unit computes while an earlier unit communicates, the two units can otherwise alias the same buffers or observe inconsistent primitive versions. Replicating model-visible state for every in-flight unit would remove aliasing but would be too expensive.

Odin avoids this overhead with \emph{Shadow Graph} (Figure~\ref{fig:shadow_graph}). \emph{Shadow Graph} gives each in-flight unit a private logical view of data access and gradient staging while keeping the underlying physical model state shared. The numbered arrows in Figure~\ref{fig:shadow_graph} have the following meaning: (1) each unit reads from a virtualized data buffer; (2) each unit executes with its own instance-local autograd graph; (3) gradients are written into a shared two-slot ring buffer that also serves as the communication source; (4) once the required primitive-publication condition is satisfied, the produced slot is committed to the shared gradient view; (5) the optimizer consumes this shared gradient view to update model data; and (6) the data view is refreshed for subsequent units. \emph{Shadow Graph} therefore resolves aliasing by virtualizing access and staging publication; this adds small metadata and staging cost, measured in Figure~\ref{fig:sdgraph}, while avoiding full model replication.

\paragraph{\textbf{Quality-first path}}
The quality-first path follows one publication rule: before a later task unit \(B\) reads mutable primitive state, every earlier unpublished update from \(A\) that could affect that read must either be disjoint under conservative scopes or already published. Let \(R^+(X)\) and \(W^+(X)\) be the conservative read and update scopes of task unit \(X\). Odin checks the primitive-state hazards
\[
\begin{aligned}
C_{\mathrm{WAW}}^{+}(A,B)&=W^{+}(B)\cap W^{+}(A),\\
C_{\mathrm{RAW}}^{+}(A,B)&=(R^{+}(B)\setminus W^{+}(B))\cap W^{+}(A),\\
C_{\mathrm{WAR}}^{+}(A,B)&=W^{+}(B)\cap (R^{+}(A)\setminus W^{+}(A)).
\end{aligned}
\]
RAW and WAW conflicts wait for \(A\)'s primitive publication; WAR conflicts wait until \(A\)'s state-observing phase has completed. The RAW expression excludes \(W^+(B)\) because overlap with \(B\)'s update scope is already covered by WAW. If all three sets are empty, \(B\) may proceed while unrelated communication remains in flight. Missing scopes, invalid structural versions, densification, pruning, reset, or overly broad mutable candidate selection synchronize the affected phase. Thus the quality-first path replaces the global barrier with primitive-level waits while preserving synchronized-training visibility under the declared primitive-scope interface.

\paragraph{\textbf{Throughput-first path}}
This path addresses a deployment pain point: in overlapping or dense PBNR captures, a small nonzero RAW overlap can still block a useful overlap window. Such overlap is not automatically high impact. Compositing, occlusion, and transmittance give primitive interactions physical and numerical weights, making some dominant and others weak~\cite{occluGS,ODAGS}. Throughput-first therefore changes only RAW waits that pass both admission tests and uses one admission parameter \(\tau\in[0,1]\) to bound delayed-scope ratio and producer-gradient magnitude. For a candidate transition, Odin defines the delayed-read set as
\[
D^{+}(A,B)=C_{\mathrm{RAW}}^{+}(A,B)
\]
and measures \(\rho(A,B)=|D^{+}(A,B)|/|R^{+}(B)|\) when \(R^+(B)\neq\emptyset\); an empty consumer read scope has no RAW delayed-read admission to perform. The delayed read is considered only when \(\rho(A,B)\le\tau\). Each delayed primitive \(p\) must also have producer-gradient evidence no larger than a \(\tau\)-scaled active-set mean, where \(\|g_A(p)\|\) is the per-primitive gradient magnitude after pipeline-level normalization:
\[
\|g_A(p)\| \le \tau \cdot
\frac{1}{|\mathcal{C}_A|}\sum_{q\in\mathcal{C}_A}\|g_A(q)\|.
\]
The gradient gate is evaluated only when \(\mathcal{C}_A\) is nonempty and all delayed primitives have producer evidence. If either test fails, or if evidence is missing, Odin waits exactly as in the quality-first path. WAW, WAR, writes, structural changes, high-impact cases, and missing evidence are never admitted by this path. Each admitted primitive reads the last version published before \(A\)'s pending update. Per-primitive publication metadata prevents a second delayed producer from being admitted on the same primitive, so a read can skip at most one unpublished update. The default \(\tau\) setting is fixed for the main results and validated in Section~\ref{sec:evaluation}; quality-sensitive reconstructions can use the quality-first path.

\paragraph{\textbf{Dynamic Data Scheduling (DDS)}}
Static Data Scheduling determines which data items belong to each task unit and in what order these units appear on the scheduled graph, but it does not fix how ready task units should be dispatched to ranks at runtime. In practice, phase durations vary across units and ranks, so purely static dispatch may expose avoidable stragglers and bubbles even when the task-unit order is well chosen.

Odin therefore performs lightweight online refinement. Whenever multiple ready task units are simultaneously admissible under current graph and synchronization rules, the runtime assigns them to ranks using recent load estimates and queue state. This refinement does not rewrite data assignment inside task units and does not alter the offline order; it only adjusts the spatial dispatch of already admissible work to reduce short-term skew and improve utilization.

\paragraph{\textbf{Dynamic Asynchronous Scheduling (DAS)}}
SAS places candidate overlap windows before exact primitive activity is known. DAS revalidates these windows as more faithful evidence becomes available. The initial RLG is constructed from stable co-visibility for static planning. Before a task unit observes mutable state, a pre-observation scope hook exposes conservative primitive scopes. After execution, the backward phase reports realized activity and gradient magnitude, refining later dispatch and throughput-first decisions.

If current scopes do not support a planned overlap, DAS inserts the missing primitive-level synchronization as a stream dependency and delays the consumer until the producer publication completes. Because this check runs before the consumer observes conflicting state, correction requires neither rollback nor kernel re-execution; it only reduces planned overlap and may expose residual bubbles. Together, DDS and DAS keep the original method symmetry: Graph Scheduling predicts data order and overlap opportunities, while Graph Execution refines dispatch and validates publication before state observation using the best available scope evidence.

\section{Implementation}
\label{sec:implementation}

Odin is implemented as a scheduling and runtime layer around existing PBNR training pipelines. The offline planner consumes camera metadata, SfM tracks, phase boundaries, and a coarse phase profile to build the RLG, form \(K\) logical groups, rotate task units, and mark candidate overlap windows. Planning is tied to stable scene metadata and training-loop phases rather than renderer internals, so it does not change model representation, renderer kernels, optimizer math, training budget, or model capacity.

At runtime, Odin requires three integration points, implemented as hooks in our adapters. A pre-state-observation scope hook exposes conservative primitive ids before mutable parameter, opacity, candidate metadata, or optimizer-state reads. A post-backward hook reports realized active primitive ids and normalized per-primitive gradient magnitudes. A publication callback marks completed communication/reduction and the corresponding optimizer update. Across 3DGS, 2DGS, TamingGS, and DashGS, adapter code maps each renderer's candidate and active primitive sets to this interface; the optimized TamingGS/DashGS kernels keep the same primitive-id namespace. If a pre-read bound is unavailable, Odin widens the scope to an owner region or synchronizes the phase. Densification, pruning, opacity reset, or optimizer/model schedule changes advance the structural version and force a barrier.

At execution time, Odin uses separate compute and communication streams, \emph{Shadow Graph} staging, structural version metadata, publication state, and active-unit metadata to realize the scheduled graph. The same publication rules apply to data-parallel and mixed-parallel execution: in DP, Odin overlaps replica communication with compatible computation; in MP, it validates sparse cross-shard exchange the same way. The implementation changes when later work may observe primitive state, while preserving the underlying PBNR training pipeline.

\section{Evaluation}
\label{sec:evaluation}

We evaluate \emph{Odin} with 8-GPU experiments across four PBNR pipelines and 13 non-city scenes, plus a separate MatrixCity mixed-parallel case study up to 64 GPUs. The experiments ask whether primitive publication improves throughput, whether throughput-first execution keeps quality within the normalized reporting band, whether quality-first falls back conservatively on dense scopes, and whether gains come from primitive-level wait hiding rather than fewer bytes or unscoped asynchrony.

\subsection{Experimental Setup}
\label{subsec:eval_setup}

\paragraph{\textbf{Hardware}}
We run on nodes with NVIDIA RTX 4090 GPUs. Within a node, GPUs use PCIe 4.0 \(\times\)16. Across nodes, we use a 160~Gbps network with eRDMA. Each node has Intel Xeon 6462C CPUs and runs Ubuntu 22.04 LTS. We evaluate both single-node and multi-node settings, scaling up to 64 GPUs.

\paragraph{\textbf{Datasets}}
We benchmark on the datasets in Table~\ref{tab:datasets}, spanning indoor, outdoor-\(360^\circ\), object-scale, and city-scale scenes. Unless noted otherwise, results are averaged over all scenes in each dataset. MatrixCity is MP-only: replicating the full city-scale tensor state and optimizer buffers for DDP exceeds per-GPU memory at the evaluated model capacity, so a DDP run would require reducing capacity or changing training.

\begin{table}[t]
\centering
\caption{Evaluation datasets. The table lists scene count, capture type, and scale; MatrixCity is a single-region MP case study, while the other datasets use both DP and MP unless stated.}
\label{tab:datasets}
\small
\setlength{\tabcolsep}{3pt}
\renewcommand{\arraystretch}{1.0}
\begin{tabular}{l c c c}
\toprule
Dataset & \#Scenes & Type & Scale \\
\midrule
MipNeRF360~\cite{mipnerf} & 9 & in/out-\(360^\circ\) & mixed \\
Tanks \& Temples~\cite{tanks} & 2 & outdoor/object & medium \\
DeepBlending~\cite{deepblending} & 2 & indoor/room & medium \\
MatrixCity~\cite{matrixcity} & 1 & aerial/city & large \\
\bottomrule
\end{tabular}
\end{table}
\FloatBarrier

\paragraph{\textbf{Pipelines and baselines}}
We evaluate four representative PBNR pipelines: 3DGS, TamingGS, DashGS, and 2DGS. For DP, we compare against PyTorch DistributedDataParallel (DDP) with the same PBNR-specific densification and pruning behavior as the corresponding pipeline; for MP, we compare against Grendel~\cite{Grendel}. Gaian uses point-based differentiable-rendering access patterns for point placement and image-to-GPU assignment. To isolate Odin's synchronization boundary, we keep placement and exchange fixed (DDP in DP, Grendel-style partitioned state in MP) and change only ordering, admission, validation, and publication. Generic distributed-training schedulers overlap communication within declared layer/tensor boundaries but do not expose primitive publication decisions; we therefore use controls for locality-only ordering, no-locality async overlap, partitionless HOGWILD-style execution~\cite{Hogwod}, and execution without dynamic validation.

All comparisons use the same renderer kernels, optimizer, densification/pruning schedule, global batch size, training iterations, communication backend, and quality checkpoint as the corresponding baseline. Odin changes task-unit ordering, admission, validation, and publication; it does not change renderer math or model capacity. Quality comparisons use the same epoch workload, random seed, and training budget, so reported deltas include any effect of Odin's ordering and execution path.

\paragraph{\textbf{Metrics}}
We report end-to-end training throughput, reconstruction quality, and runtime overhead. Throughput is
\[
\text{throughput}=\frac{N_{\mathrm{iter}}\cdot B}{T_{\mathrm{wall}}},
\]
where \(N_{\mathrm{iter}}\) is the number of iterations, \(B\) is global batch size, and \(T_{\mathrm{wall}}\) is wall-clock time. Unless otherwise stated, the per-GPU batch size is \(B_{\mathrm{gpu}}=2\). Reconstruction quality uses PSNR, SSIM, and LPIPS~\cite{LPIPS}; quality figures report normalized aggregate deltas from the synchronized baseline against a \(\pm1\%\) reporting band. We also report communication hiding, overheads, and fallback behavior.

\paragraph{\textbf{Execution paths}}
We report a quality-first path and a throughput-first path. The quality-first path admits only disjoint or already published primitive scopes and conservatively synchronizes every conflicting primitive update. The throughput-first path enables importance-aware admission with logical-region count \(K=4\) and joint admission parameter \(\tau=0.2\) unless stated otherwise; \(\tau\) bounds the delayed primitive-set ratio and selects gradients below a \(\tau\)-scaled active-set mean. The headline \(1.22\times\) is the arithmetic mean of Odin/base throughput ratios over the evaluated non-city 8-GPU runs with throughput-first execution; Figure~\ref{fig:model} groups the same runs for readability, so visible group labels are not an averaging recipe. We also report the quality-first path on an 8-GPU 3DGS fallback stress subset; with importance-aware delayed reads disabled, speedup ranges from \(1.00\times\) on dense \textit{kitchen} to \(1.23\times\) on sparse scenes.

\subsection{End-to-End Acceleration}
\label{subsec:eval_perf}

\begin{figure}[!htbp]
    \centering
    \includegraphics[width=0.98\columnwidth]{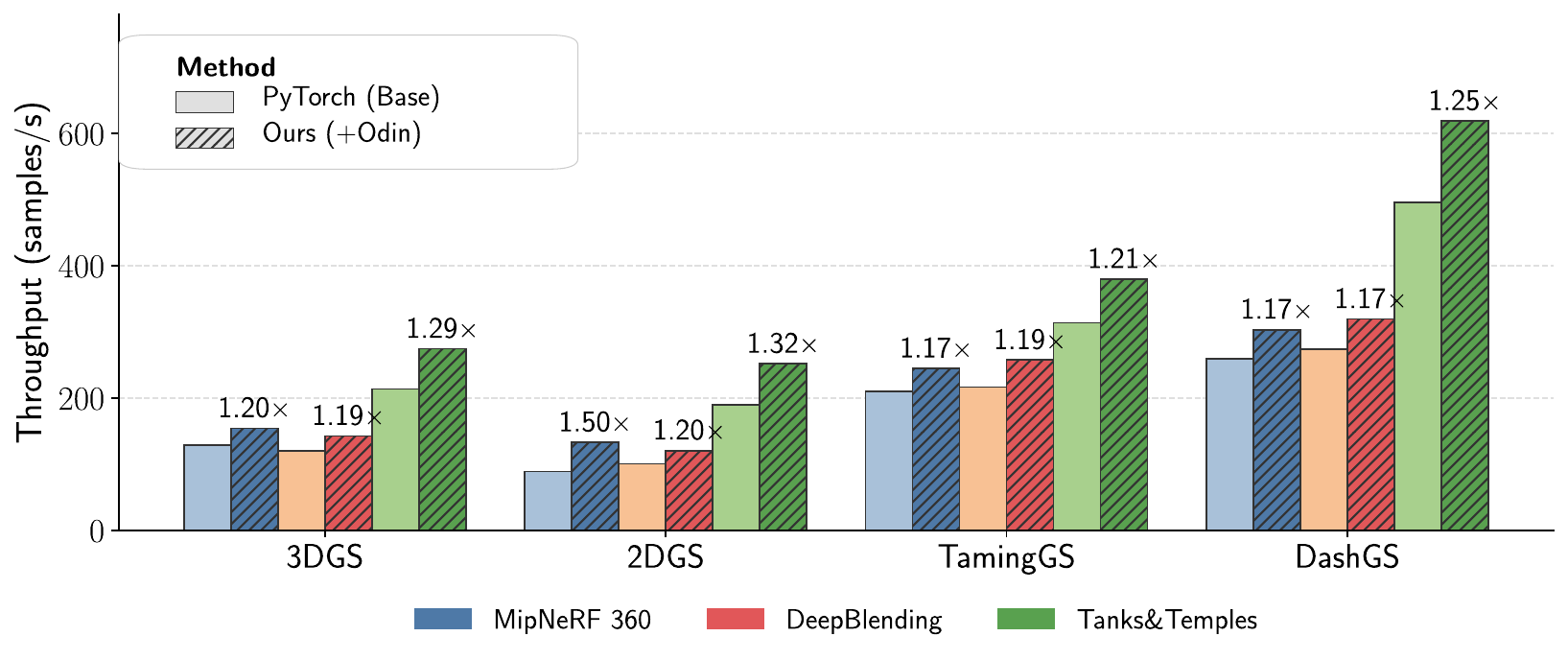}
    \caption{End-to-end 8-GPU throughput, excluding MatrixCity. Each group is one pipeline--dataset setting; paired bars compare the task- or iteration-synchronized baseline with Odin's default throughput-first setting, labels are grouped speedup, and higher samples/s is better. The headline \(1.22\times\) is computed over scene-level non-city 8-GPU Odin/base ratios.}
    \label{fig:model}
\end{figure}

\begin{figure}[!htbp]
    \centering
    \includegraphics[width=0.96\columnwidth]{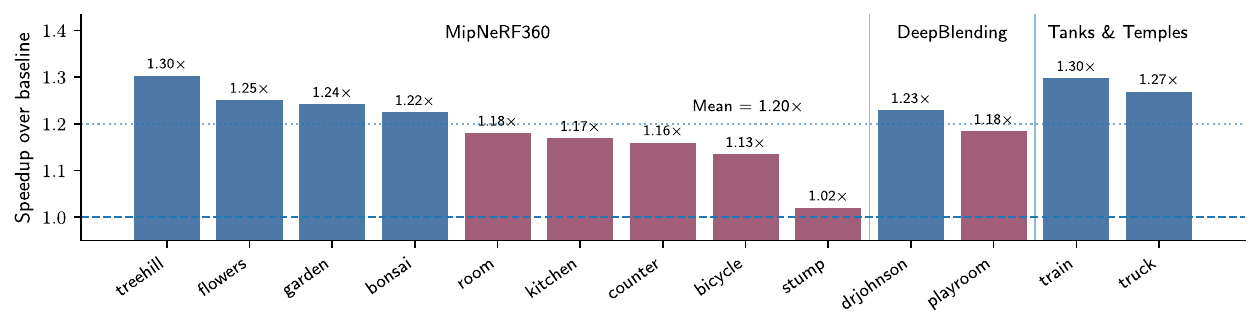}
    \caption{Per-scene 3DGS speedup over DDP on 8 GPUs. One bar is one scene; colors distinguish above-mean and below-mean gains. Higher observed speedup corresponds to more admitted overlap under this setup. \textit{Stump} is the low-overlap case.}
    \label{fig:scene}
\end{figure}

\begin{figure}[!htbp]
    \centering
    \includegraphics[width=0.98\columnwidth]{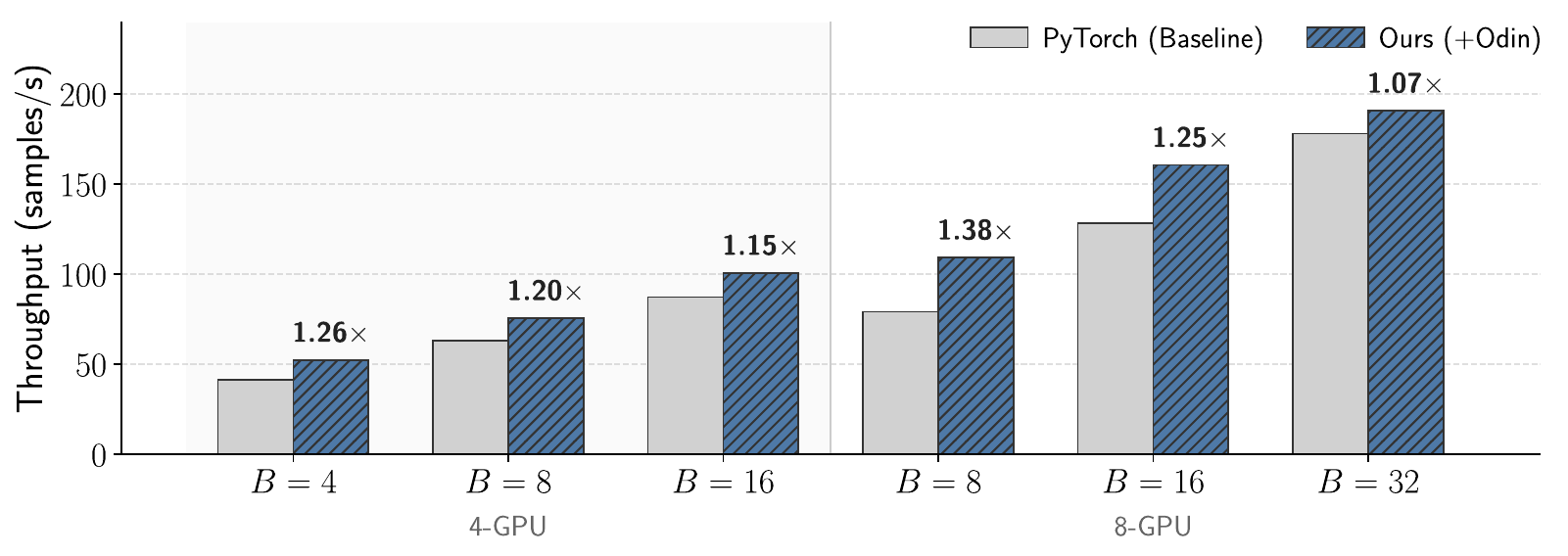}
    \caption{3DGS throughput on \textit{DeepBlending} across GPU counts and global batch \(B\). Gray is PyTorch/DDP, blue is Odin, labels are speedup over DDP, and higher samples/s is better. Larger \(B\) narrows relative gain.}
    \label{fig:speedbatch}
\end{figure}

\paragraph{\textbf{Single-node DP}}
Figures~\ref{fig:model}--\ref{fig:speedbatch} test single-node DP training with the same renderer kernels and communication backend as the baseline. Figure~\ref{fig:model} shows gains across 3DGS, 2DGS, TamingGS, and DashGS because all retain iteration-level publication. Gains are larger on less optimized pipelines, where communication and rank skew remain more exposed; TamingGS and DashGS leave less removable waiting after their kernel and update-pipeline optimizations.

\noindent
Figure~\ref{fig:scene} breaks down the same question by scene. Per-scene speedup varies substantially. The low-gain \textit{Stump} case is consistent with denser view--primitive coupling and fewer admitted overlap windows, while higher-gain scenes leave cleaner structural signals and more decomposable interaction patterns. Speedup therefore follows dependency structure rather than pipeline identity alone.

\noindent
Figure~\ref{fig:speedbatch} sweeps GPU count and global batch size. Odin improves throughput throughout the sweep, while relative gain narrows as \(B\) increases because larger batches create fewer synchronization boundaries per sample. This is the expected signature of a primitive-level synchronization optimization: it helps most when communication remains on the critical path and enough later work exists to overlap with it.

\begin{figure}[!htbp]
    \centering
    \includegraphics[width=0.96\columnwidth]{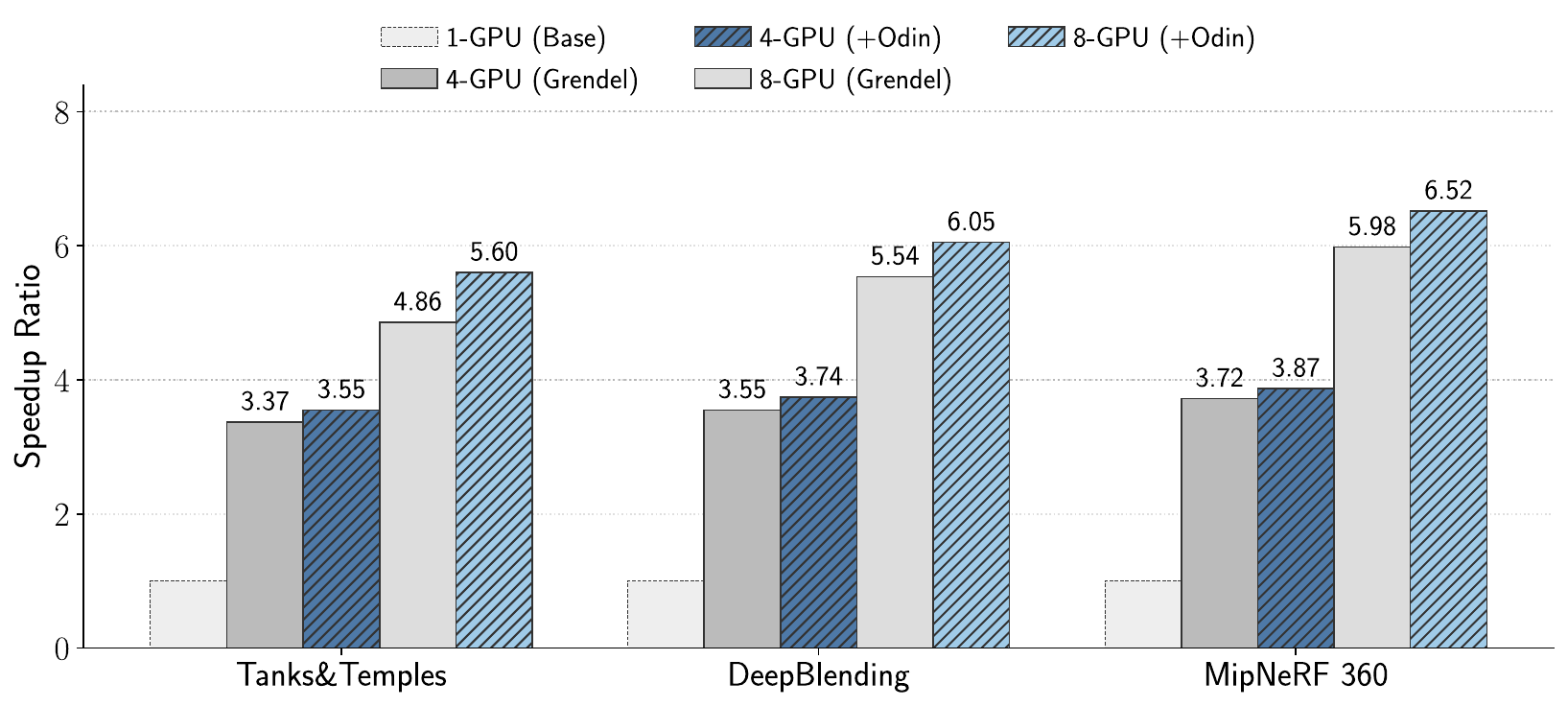}
    \caption{Mixed-parallel scaling across datasets. Dashed line is the 1-GPU reference; bars compare Grendel with Grendel+Odin in each regime; labels are overall speedup over the 1-GPU reference, and higher speedup is better.}
    \label{fig:speed}
\end{figure}

\begin{figure}[!htbp]
    \centering
    \includegraphics[width=0.96\columnwidth]{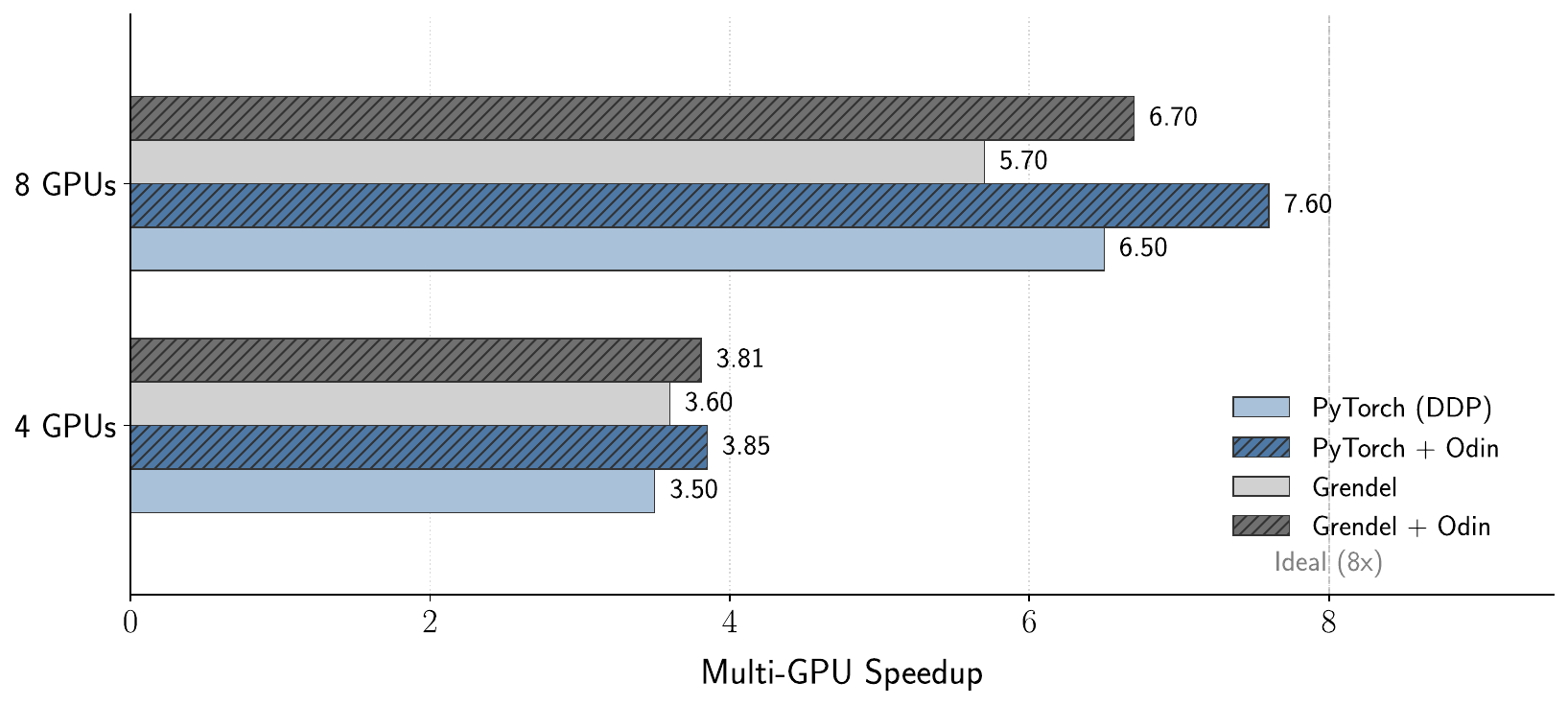}
    \caption{Overall multi-GPU speedup on \textit{truck}. Bars compare PyTorch/DDP, PyTorch+Odin, Grendel, and Grendel+Odin against the 1-GPU reference; the marker is ideal 8-GPU speedup. Odin reaches \(7.6\times\).}
    \label{fig:speed2}
\end{figure}

\begin{figure}[!htbp]
    \centering
    \includegraphics[width=0.96\columnwidth]{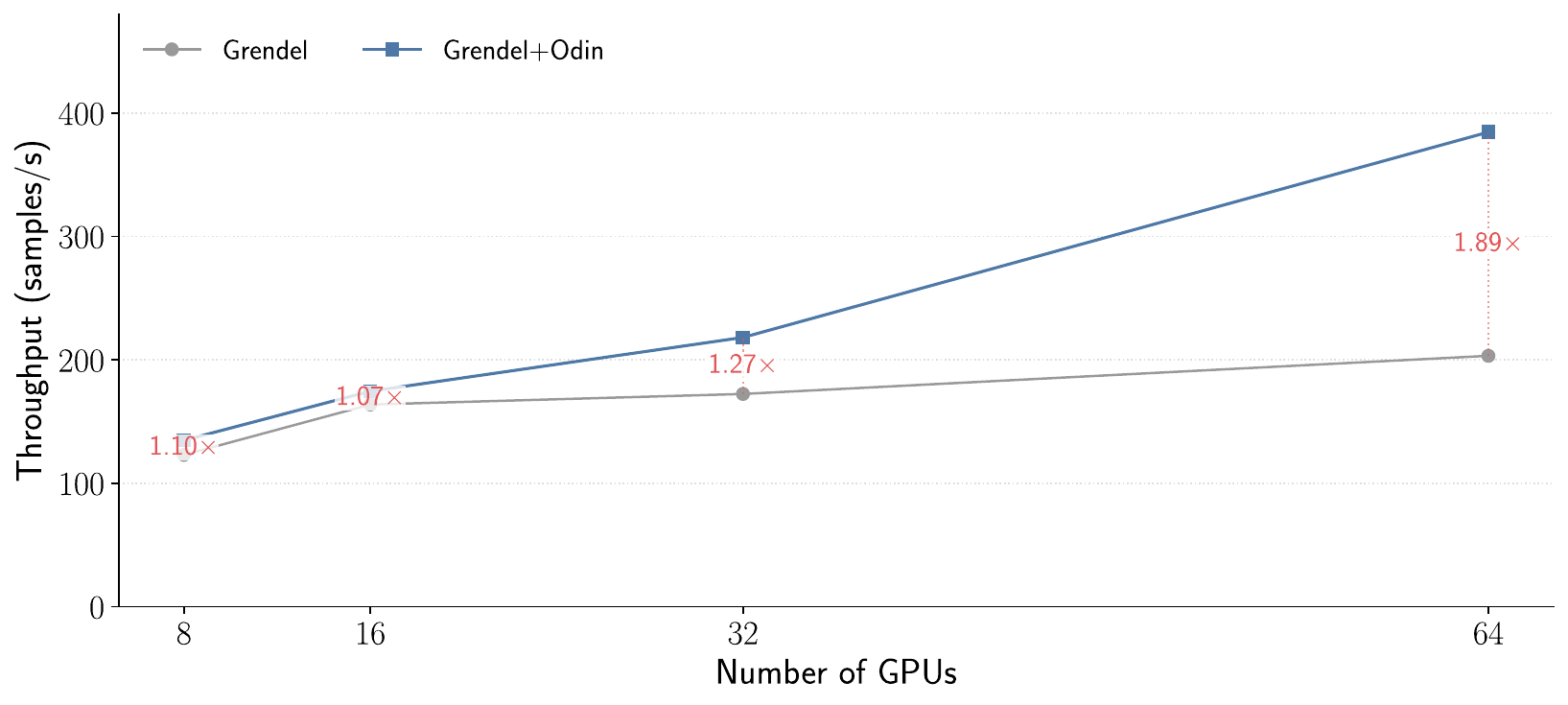}
    \caption{Multi-node MP throughput on single-region MatrixCity. X-axis is GPU count, y-axis samples/s, gray Grendel, blue Grendel+Odin, and red labels Odin-over-Grendel speedup: \(1.27\times\) at 32 GPUs and \(1.89\times\) at 64 GPUs.}
    \label{fig:scaling}
\end{figure}

\paragraph{\textbf{Parallel regimes and scale}}
Figures~\ref{fig:speed}--\ref{fig:scaling} test whether Odin remains useful when the baseline already uses partitioned state. Grendel reduces memory pressure and communication volume through spatial partitioning; Odin still improves throughput because it addresses whether the remaining sparse exchange must block unrelated later work. The gain is smaller than the DP gain at some single-node points because Grendel already removes part of the exposed cost, but it persists in the fragmented MP regime. Primitive-level synchronization is therefore complementary to communication sparsification.

\noindent
Figure~\ref{fig:speed2} compares DP and MP on \textit{truck}. Adding Odin moves both regimes closer to ideal scaling, reaching \(7.6\times\) with DDP+Odin and improving Grendel as well. Figure~\ref{fig:scaling} extends the result to multi-node MatrixCity: gain grows from \(1.27\times\) at 32 GPUs to \(1.89\times\) at 64 GPUs, consistent with greater synchronization exposure at larger scale. Even in this communication-heavy case, Odin improves effective scaling by moving compatible sparse exchange off the critical path.

\subsection{Operating Point and Overheads}
\label{subsec:eval_tradeoff}

\begin{figure}[!htbp]
    \centering
    \includegraphics[width=0.96\columnwidth]{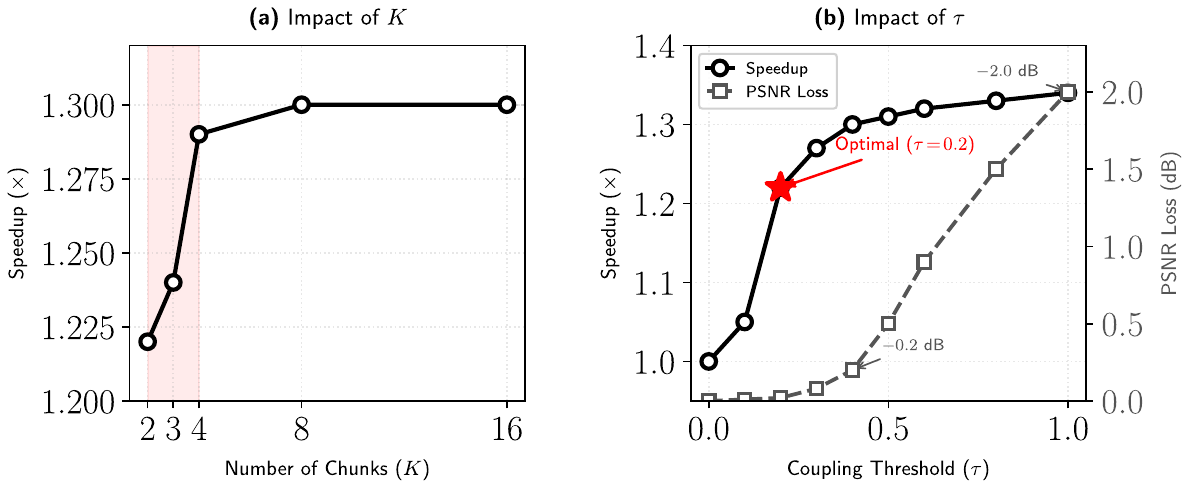}
    \caption{Scheduling-parameter sensitivity. Left sweeps logical-region count \(K\); right sweeps joint admission parameter \(\tau\). Default setting: \(K=4,\tau=0.2\).}
    \label{fig:ktau}
\end{figure}

\begin{figure}[!htbp]
    \centering
    \includegraphics[width=0.96\columnwidth]{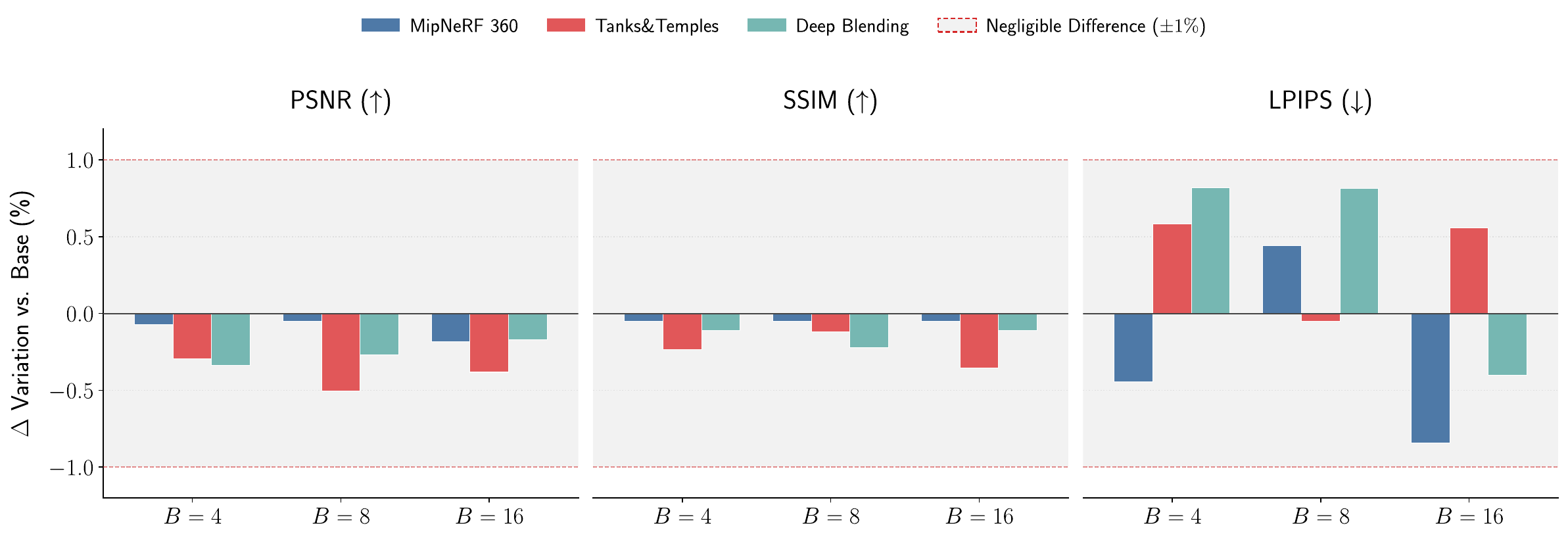}
    \caption{Reconstruction quality at \(\tau=0.2\). Groups sweep global batch \(B\), colors are datasets, and bars are normalized aggregate deltas from the task- or iteration-synchronized baseline. Dashed box marks the \(\pm1\%\) reporting band.}
    \label{fig:quality}
\end{figure}

\begin{figure}[!htbp]
    \centering
    \includegraphics[width=0.96\columnwidth]{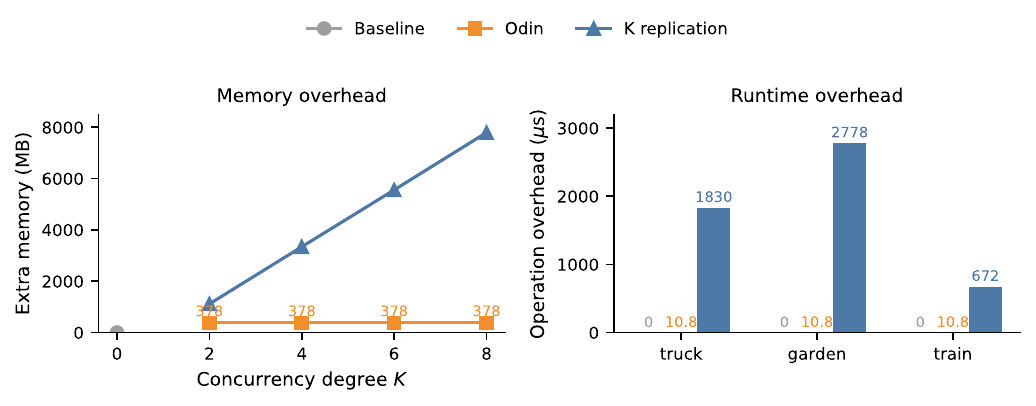}
    \caption{\emph{Shadow Graph} overhead. Left: extra memory versus logical-region count \(K\); right: per-operation virtualization cost. Compared with full logical-state replication, Odin keeps memory nearly flat and cuts operation cost by up to \(257\times\).}
    \label{fig:sdgraph}
\end{figure}

\begin{figure}[!htbp]
    \centering
    \includegraphics[width=0.96\columnwidth]{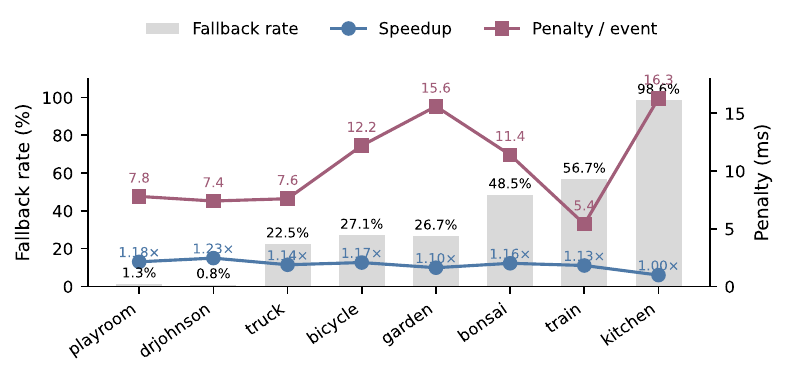}
    \caption{Runtime fallback under the quality-first path. Gray bars are rejected planned overlaps, blue points speedup, and purple points wait per fallback event. Sparse \textit{Playroom}/\textit{drjohnson} still speed up; dense \textit{kitchen} falls back to synchronization.}
    \label{fig:fallback_overhead}
\end{figure}

Figures~\ref{fig:ktau}--\ref{fig:fallback_overhead} and Table~\ref{tab:mapping_overhead} characterize throughput-first sensitivity, quality, ahead-of-time cost, \emph{Shadow Graph} overhead, and quality-first fallback.

\noindent
Figure~\ref{fig:ktau} shows that increasing \(K\) or \(\tau\) can enlarge the schedulable region, but the return is not monotonic. Too few logical regions leave coupled views adjacent; too many fragment scheduling and increase live-state pressure. A larger \(\tau\) admits more pending updates by widening both tests. The default setting \(K=4,\tau=0.2\) is used unchanged in the main results.

\noindent
Figure~\ref{fig:quality} evaluates the default throughput-first setting. At \(K=4,\tau=0.2\), all plotted PSNR, SSIM, and LPIPS aggregate deltas stay within the \(\pm1\%\) reporting band. This matches the admission rule: delayed reads are limited to small nonzero overlap scopes and weak producer-gradient updates, which typically correspond to non-dominant interactions under PBNR compositing and occlusion. The quality-first path remains the conservative choice when delayed reads are unacceptable.

\begin{table}[h]
\centering
\caption{End-to-end ahead-of-time (AOT) overhead from track metadata, in seconds. Includes graph construction and schedule compilation; excludes SfM/reconstruction.}
\label{tab:mapping_overhead}
\scriptsize
\setlength{\tabcolsep}{4pt}
\begin{tabular}{@{}lcccccc@{}}
\toprule
\textbf{Images} & 1,000 & 2,000 & 5,000 & 10,000 & 20,000 & 50,000 \\
\midrule
\textbf{Time} & 0.013 & 0.028 & 0.161 & 0.464 & 1.673 & 9.450 \\
\bottomrule
\end{tabular}
\end{table}

\noindent
Table~\ref{tab:mapping_overhead} shows that the ahead-of-time scheduler is lightweight for production-scale captures: 50{,}000 images take 9.45 seconds for graph construction and schedule compilation, excluding SfM/reconstruction already present in common PBNR inputs. The dominant work is the inverse track-to-image mapping; scheduling itself adds little overhead. This one-time cost is amortized over training.

\noindent
Figure~\ref{fig:sdgraph} shows that \emph{Shadow Graph} avoids full logical-state replication: extra memory stays nearly flat with \(K\), and per-operation overhead is up to \(257\times\) lower than the replication-based alternative.

\noindent
Figure~\ref{fig:fallback_overhead} disables importance-aware delayed reads to stress quality-first execution. Sparse \textit{playroom}/\textit{drjohnson} reject only 1.3\% and 0.8\% of planned overlaps, while dense \textit{kitchen} rejects 98.6\% and returns to \(1.00\times\). Intermediate scenes retain speedup despite nontrivial fallback; failed predictions become primitive-scoped synchronization rather than rollback or re-execution. The pattern matches Odin's conservative fallback design.

\subsection{Ablation Study}
\label{subsec:eval_ablation}

\begin{figure}[!htbp]
    \centering
    \includegraphics[width=0.96\columnwidth]{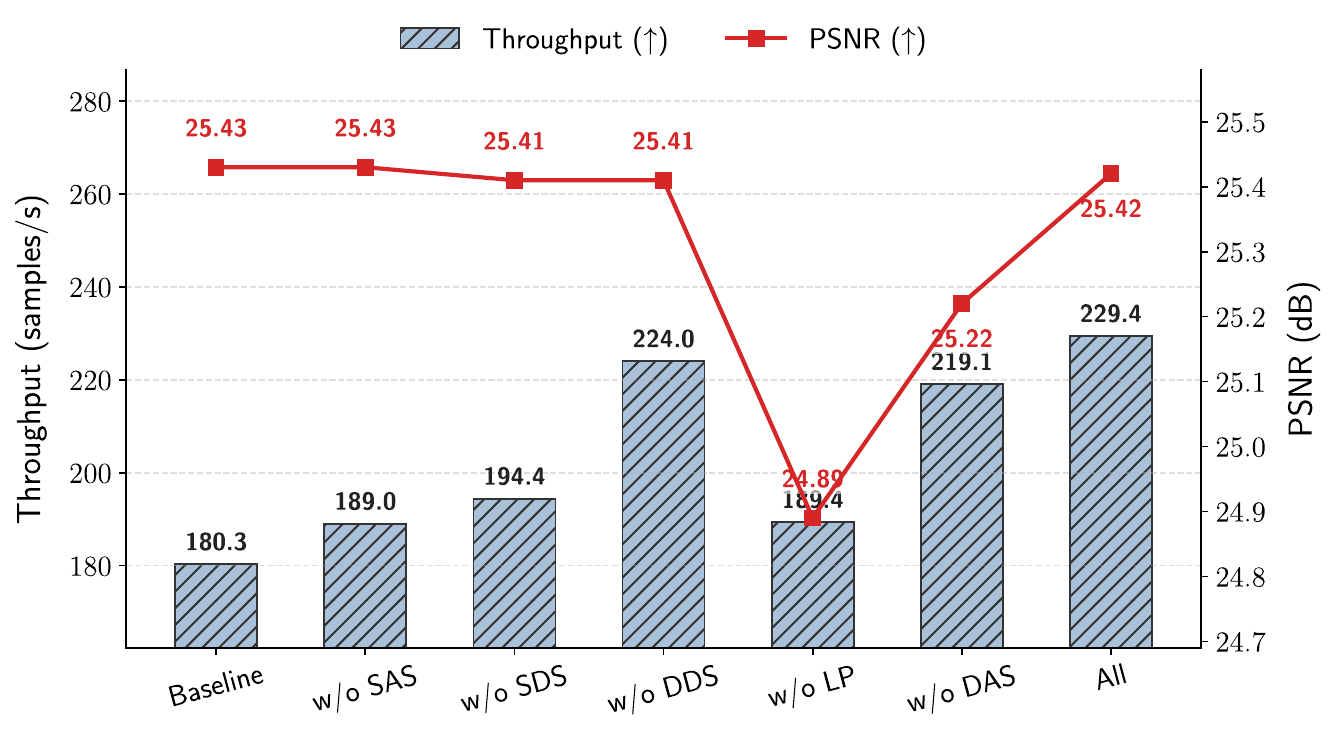}
    \caption{Ablation on a representative 8-GPU 3DGS run (\(K=4,\tau=0.2\)). Bars show throughput, red line shows PSNR, and ``w/o LP'' is the partitionless HOGWILD-style control.}
    \label{fig:ablation}
\end{figure}

\begin{figure}[!htbp]
    \centering
    \includegraphics[width=0.96\columnwidth]{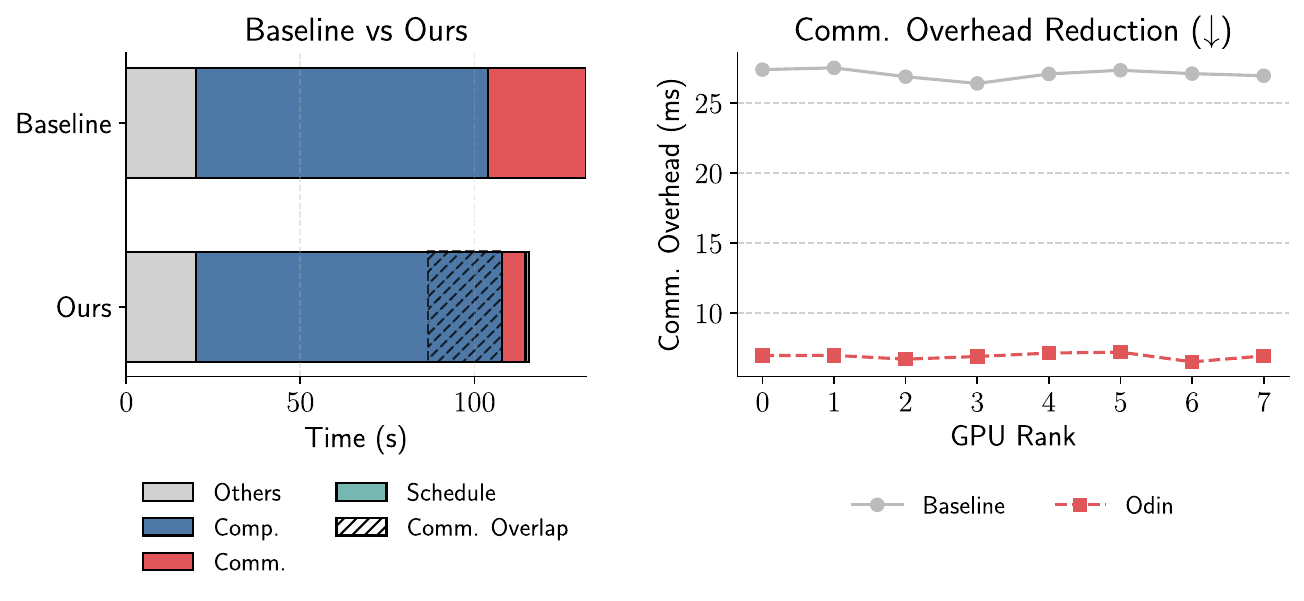}
    \caption{Runtime breakdown on a representative 8-GPU 3DGS run. Left shows timeline components; right shows exposed wait after overlap. Odin hides 82\% of critical-path wait, not raw bytes.}
    \label{fig:ratio}
\end{figure}

\paragraph{\textbf{Component and wait attribution}}
Figures~\ref{fig:ablation} and~\ref{fig:ratio} attribute gains to coordinated ordering, admission, validation, and wait hiding.

\noindent
Figure~\ref{fig:ablation} shows that removing SAS largely removes the gain, confirming that barrier removal and communication--computation overlap are the main source of acceleration. Removing SDS also hurts because useful overlap depends on arranging weakly coupled transitions near candidate windows. The dynamic components are complementary: DDS reduces short-term imbalance, while DAS protects the plan from outdated or inaccurate locality predictions. The partitionless asynchronous control loses throughput and quality, arguing against a simple HOGWILD-style interpretation. Full Odin is therefore a coordinated design rather than a single rule.

\noindent
Figure~\ref{fig:ratio} connects the ablation back to the motivation: Odin hides 82\% of critical-path exposed wait through primitive-level synchronization and validation, not renderer changes, model reduction, or unscoped asynchrony. The remaining exposed time comes from globally synchronized phases, rejected overlap windows, or ranks without enough compatible ready work.

\section{Related Work}
\label{sec:related}

\paragraph{\textbf{PBNR representation and single-device systems}}
Recent PBNR work improves scene representation or single-device execution. Representation methods revise primitive geometry, compactness, or quality~\cite{2DGS, sugar}, while system optimizations accelerate kernels, memory movement, pruning, or update pipelines~\cite{TamingGS, DashGS, GScore, GSArch, speedysplat, 3dgslm, clm, GSScale, MetaSapiens}. These advances are complementary: they shrink computation and expose the barrier mismatch that Odin targets. Odin changes when primitive updates become visible to later work, not the representation or renderer kernels.

\paragraph{\textbf{Distributed PBNR systems}}
Existing distributed PBNR systems mainly optimize where state and work reside. CityGaussian, VastGaussian, hierarchical 3DGS, and DOGS use spatial decomposition, hierarchy, or blockwise consensus to reduce memory pressure and communication~\cite{cityGS, vastgs, h3D, Dogs}; Grendel partitions parameters and uses mixed parallelism~\cite{Grendel}. Gaian~\cite{zhao2025scalingpointbaseddifferentiablerendering} is the closest work in spirit: it exploits point-based differentiable-rendering access patterns for locality-aware point placement and image-to-GPU assignment. Odin uses the same broad property--localized primitive access--for a different boundary. Placement decides where primitives and views run and how much state moves; Odin decides when a pending primitive update must become visible at a later state read.

\paragraph{\textbf{General distributed training and relaxed updates}}
General distributed-training systems hide communication through collective scheduling, layer or tensor overlap, or pipeline stages~\cite{coconet, DeAR, Concerto, gpipe, bytescheduler}. These techniques assume explicit neural-network boundaries such as layers, tensors, or stages. PBNR has a shallow phase structure and a mutable primitive set whose active scope depends on view-dependent numerical state, so the relevant conflict boundary lies below the exposed abstraction. Delayed-update and partial-update methods~\cite{Preduce, eagarSGD, sunco2, Hogwod} study stale or partial updates in dense neural networks, sparse optimization, or tensor averaging, but they cannot be directly applied to PBNR without identifying low-impact primitive RAW overlaps. Odin specializes this case with static locality-guided scheduling, runtime primitive-scope validation, and a throughput path gated by delayed-scope ratio and producer-gradient evidence.

\section{Discussion}
\label{sec:discussion}

\paragraph{\textbf{Scope and fallback}}
Odin requires conservative primitive scopes before state observation, publication points after communication/update, and late active-id and gradient evidence. If these signals are unavailable, or if structural mutations, dense fields, global regularizers, or unversioned mutable candidate selection dominate, Odin widens the scope or synchronizes the affected phase.

\paragraph{\textbf{Relationship to other scaling techniques}}
Odin is not a replacement for partitioning, sparse exchange, or renderer optimization. Gaian and Grendel decide where primitives and views run; Odin decides when pending primitive updates become visible at state reads.

\paragraph{\textbf{Operating modes}}
Quality-first is the conservative choice for dense scenes, strict quality constraints, or insufficient scope evidence. Throughput-first targets latency-sensitive runs and admits only gated RAW delayed reads; rejected windows fall back to primitive synchronization.

\paragraph{\textbf{Beyond PBNR}}
Odin targets a workload class rather than a single application: explicit mutable state with read/write scopes, publication points, and late refinement evidence. This condition explains why the same mechanism applies across 3DGS, 2DGS, TamingGS, DashGS, DP, and MP without changing renderer kernels or optimizer math. It also suggests a broader direction for sparse explicit-state AI systems such as online reconstruction, neural mapping, and object-, voxel-, or map-level world models; dense all-to-all state, hidden mutable kernel state, or unavoidable global regularization remains conservative.

\section{Conclusion}
\label{sec:conclusion}

Odin shows that global barriers are not inherent to distributed PBNR. Primitive-level publication, static locality planning, runtime validation, and \emph{Shadow Graph} staging improve throughput by 1.22\(\times\) on average and 1.89\(\times\) over Grendel without changing kernels, optimizers, budgets, or model capacity.

\bibliographystyle{ACM-Reference-Format}
\bibliography{section/ref2}

\end{document}